\documentclass[]{JHEP3}

\usepackage{amsmath,amssymb,amsbsy}
\usepackage[all]{xy}         
\usepackage{booktabs}        

\newcommand{\N}{\mathcal{N}}
\newcommand{\hN}{\hat{N}}

\newcommand{\ped}[1]{_{\mbox{\scriptsize{#1}}}}
\newcommand{\api}[1]{^{\mbox{\scriptsize{#1}}}}

\newcommand{\llangle}{\langle \hspace{-0.05em} \langle}
\newcommand{\rrangle}{\rangle \hspace{-0.05em} \rangle}

\newcommand{\one}{\hbox{1\kern-.8mm l}}
\DeclareMathOperator{\e}{e}
\DeclareMathOperator{\tr}{tr}
\DeclareMathOperator{\Tr}{Tr}

\newcommand*{\du}{\mathop{}\!\mathrm{d}}
\newcommand*{\diag}{\mathop{}\!\mathrm{diag}}
\newcommand{\sun}[1]{\textrm{SU}(#1)}
\newcommand{\un}[1]{\textrm{U}(#1)}
\newcommand{\um}{\frac{1}{2}}

\title{Matrix Model and $\beta$-deformed ${\cal N} =4$ SYM }

\author{G.C.~Rossi$^{ab}$, M.~Siccardi$^{cd}$, Ya.S.~Stanev$^{b}$ and K.~Yoshida$^{cd}$ \\ $^a$ Dipartimento di Fisica, Universit\`a di Roma ``{\it Tor Vergata}'' \\ Via della Ricerca Scientifica - 00133 Roma, Italy \\ $^b$ INFN, Sezione di Roma ``{\it Tor Vergata}''\\ Via della Ricerca Scientifica - 00133 Roma, Italy \\ $^c$ Dipartimento di Fisica, Universit\`a di  Roma ``{\it La Sapienza}'' \\ Piaz.le A. Moro 2 - 00185 Roma, Italy \\ $^d$ INFN, Sezione di Roma ``{\it La Sapienza}'' \\ Piaz.le A. Moro 2 - 00185 Roma, Italy \\ E-mail: \email{giancarlo.rossi@roma2.infn.it}, \email{matteo.siccardi@roma1.infn.it}, \email{yassen.stanev@roma2.infn.it}}

\preprint{ROM2F/2009/18}

\abstract{This work is the result of the ideas developed by Ken Yoshida about the possibility of extending the range of applications of the matrix model approach to the computation of the holomorphic superpotential of the $\beta$-deformed ${\cal N} =4$ super Yang-Mills theory both in the presence of a mass term and in the massless limit. Our formulae, while agreeing with all the existing results we can compare with, are valid also in the case of spontaneously broken gauge symmetry.}

\dedicated{We dedicate this paper to the memory of Ken,\\ an unforgettable friend for all of us\\ and a great scientist.}

\keywords{Supersymmetric Gauge Theories, Matrix Models, Supersymmetric Effective Theories}

\begin{document}

\section{Introduction}

Understanding the precise infrared behaviour of asymptotically free gauge theories is a daunting task. For instance, determination of the confining properties of the theory has proven to be impossibly difficult. The reason for the difficulty in analysing the low-energy dynamics of the theory is related to the fact that perturbative methods have a limited range of applicability as the coupling is growing larger in the infrared. On the other hand, we lack systematic non-perturbative methods. Up to date, the most powerful tool at our disposal is undoubtedly represented by lattice simulations. Still this is a numerical approach and any possible piece of analytic information would be greatly valued.

Interestingly, however, there exist relevant exceptions to the previous picture represented by supersymmetric theories. The constraints on the structure of the theory imposed by supersymmetry are so strong that it is possible to obtain detailed low energy information about holomorphic quantities, such as the holomorphic superpotential. Holomorphicity, in fact, protects physical quantities from quantum corrections and/or allows to safely control them providing numerous and very useful non-renormalization theorems~\cite{Piguet:1986td,Seiberg:1988ur,Seiberg:1994bz}.

In recent years, there has been a renewed interest in the field after Dijkgraaf and Vafa conjectured that holomorphic quantities in the low-energy regime are related to the free-energy of an hermitian Matrix Model (MM). The correspondence has passed various non-trivial tests~\cite{Cachazo:2002ry,Cachazo:2003yc} and it has become a powerful tool for studying the low-energy limit of $\N=1$ gauge theories with massive chiral superfields.

At the same time, a particular $\N=1$ model, known as the  $\beta$-deformation of the $\N=4$ Super Yang-Mills ($\beta$d SYM), attracted the attention of the community. Leigh and Strassler discovered this model in the nineties~\cite{Leigh:1995ep} and various aspects of the model have been studied~\cite{Berenstein:2000ux,Aharony:2002hx,Benini:2004nn}, but only after its gravity dual was found~\cite{Lunin:2005jy} the $\beta$d SYM was extensively investigated in numerous papers~\cite{Frolov:2005ty,Mauri:2005pa,Rossi:2006mu,Ananth:2006ac,Mansson:2007sh,Kazakov:2007dy}.

In this paper we apply the Dijkgraaf--Vafa correspondence to the study of the holomorphic low-energy superpotential of the $\beta$d SYM model. In particular, we are able to compute the low-energy coupling constants controlling the surviving massless degrees of freedom, after the $\un{N_c} \mapsto \prod_{i=1}^n \un{N_i}$ spontaneous symmetry-breaking has occurred. Along the way, as a side result, we also compute the low-energy effective superpotential of the massive version of the $\beta$d SYM model as an inverse mass power expansion.

The paper is organized as follows. In section~\ref{sec:PM} we review the standard lore about the MM approach and the general properties of various deformations of the $\N=4$ SYM model. In section~\ref{sec:computation1}, relying on MM techniques, we compute the holomorphic low energy superpotential in various situations. In section~\ref{sec:computation2} we investigate the phenomenon of spontaneous breaking of the gauge symmetry. We end in section~\ref{sec:conclusions} with some conclusion and a few considerations about possible lines of future developments. A few more technical issues are discussed in appendices.

\section{Preliminary material}\label{sec:PM}

\subsection{The Matrix Model}

It has been suggested~\cite{Dijkgraaf:2002dh} that for a wide class of supersymmetric gauge field theories there exists a deep connection between the ``free-energy'' of certain (zero-dimensional) MM's and the holomorphic superpotential as function of the glueball superfield degrees of freedom.

In its original formulation it was believed that such a correspondence was limited to the \textit{perturbative} corrections to the low-energy superpotential, thus explicitly excluding the celebrated $\N=1$ Veneziano--Yankielowicz (VY) superpotential~\cite{Veneziano:1982ah,Cachazo:2002ry,Argurio:2003ym}, $W\ped{VY}$. This deficiency was attributed to the fact that the overall constant of the matrix model integral measure could not be unambiguously fixed. In~\cite{Ooguri:2002gx} a first attempt was made to relate it to the gauge fixing procedure necessary to give meaning to the matrix model\footnote{A similar idea was brought up, in the context of the Eguchi--Kawai one-plaquette model~\cite{Eguchi:1982nm}, in ref.~\cite{Rossi:1982zg}.}, but a clear cut derivation of $W_{\rm VY}$ was not provided.

In this context the works of refs.~\cite{Kawai:2003yf,Kawai:2003dw} appeared, where the direct correspondence between the Dijkgraaf--Vafa approach and certain generalizations of gauge field theories (on non-commutative space-time) was displayed. Immediately after, in ref.~\cite{Kawai:2004bz}, the boundary condition constraint imposed by the triviality of the $\N=4$ superpotential was exploited to determine the unknown overall coefficient of the MM functional integral, much in the spirit of~\cite{Arnone:2004ek} thus allowing the determination of $W_{\rm VY}$.

For the sake of completeness we want to review the derivation of ref.~\cite{Kawai:2004bz}. There one starts with a $\hN$-dimensional MM characterized by the tree-level action (potential) for the hermitian $\hN\times\hN$ matrices $\hat{\Phi}_I\, , I=1,2,3$
\begin{equation}\label{eq:mm_sm}
S_m=\frac{\hN}{g_m} \tr \bigg( g \hat{\Phi}_1 [\hat{\Phi}_2,\hat{\Phi}_3]+W\ped{aux}(\hat{\Phi}_1)+\frac{M_2}{2} \hat{\Phi}_2^2+\frac{M_3}{2}\hat{\Phi}_3^2\bigg)
\end{equation}
and the definition of the Dijkgraaf--Vafa-type free-energy
\begin{equation}
Z_m = \exp \left[ - \frac{\hN^2}{g_m^2} F_m \right]=C_{\hN} \int \du \hat{\Phi}_1 \du \hat{\Phi}_2 \du \hat{\Phi}_3 \exp[-S_m].
\label{eq:ZETA}
\end{equation}
In the above formulae, $g_m$ only plays the r\^{o}le of a scaling constant for the matrix action but in the following we will uncover its meaning on the gauge theory side. The matrix function $W\ped{aux}(\hat{\Phi}_1)$ is an auxiliary potential term chosen according to computational needs. For instance, to determine the overall normalization coefficient $C_{\hN}$, one has to consider the specific form $W\ped{aux}(\hat{\Phi}_1)=\um M_1 \hat{\Phi}_1^2$ leading to
\begin{equation}
S_m=\frac{\hN}{g_m} \tr \bigg( g\hat{\Phi}_1[\hat{\Phi}_2,\hat{\Phi}_3]+\frac{M_1}{2}\hat{\Phi}_1^2 + \frac{M_2}{2}\hat{\Phi}_2^2+\frac{M_3}{2}\hat{\Phi}_3^2\bigg).
\end{equation}
This is nothing else but the so-called MM formulation of the $\N=1^*$ model. The $\N=1^*$ model is the $\N=4$ SYM theory supplemented by mass terms for the chiral superfields (all belonging to the adjoint representation of the gauge group $\un{N_c}$). For large values of the mass parameters $\{M_I\} \equiv \{M_1, M_2, M_3\}$, this model can be regarded as a regularization of the $\N=1$ SYM theory by means of a mass deformation of $\N=4$ SYM~\cite{Arkani-Hamed:1997mj,Arnone:1998zc}. The $\N=1^*$ model is free of UV divergences for arbitrary masses just like the original $\N=4$ (superconformal) theory~\cite{Howe:1982tm,Howe:1983sr,Mandelstam:1982cb,Kovacs:1999fx,Kovacs:1999rd}, as no new UV divergences arise upon introducing mass terms.

The pure $\N=1$ SYM is reached in the $\{M_I\} \equiv M_0 \to \infty$ and $g \to 0$ limit keeping fixed
\begin{equation}\label{eq:Lambda}
  \Lambda_{\N=1} = M_0 \exp \left( - \frac{8 \pi^2}{3 N_c g^2}\right),
\end{equation}
while in the $\{M_I\} \to 0$ limit, the $\N=4$ SYM theory is recovered for any fixed value of $g$.

The key observation in the determination of $C_{\hN}$ is that in the $\{M_I\} \to 0$ limit (where $\N=4$ SYM is reobtained) one should find that the only holomorphic contribution to the superpotential is just given by the tree-level kinematical term. Dijkgraaf and Vafa instruct us to compute the effective superpotential by taking the derivative of the \textit{planar} (which is the contribution singled out in the $\hN \to \infty$ limit) MM free-energy with respect to $\mathcal{S}$, once the latter is introduced in place of the MM coupling constant $g_m$. In formulae,
\begin{equation}
  W\ped{eff} \equiv N_c \frac{\partial F_m (g_m \rightarrow \mathcal{S})}{\partial \mathcal{S}}, \qquad F_m \equiv - \lim_{\hN \to \infty} \frac{g_m^2}{\hN^2} \log Z_m.
\label{eq:FreeE}
\end{equation}
Then, $C_{\hN}$ is found requiring that the $\{M_I\} \to 0$ limit of the $\N = 1^*$ model in the MM setup gives for the free-energy the result
\begin{equation}\label{eq:FTRI}
  F_m^{\N=4}=\lim_{\{M_I\} \to 0} F_m^{\N=1^*} = \frac{\pi\imath\tau_0 g_m^2}{N_c} \qquad \Rightarrow \qquad W\ped{eff} = 2 \pi \imath \tau_0 \mathcal{S},
\end{equation}
where we have introduced the complexified gauge coupling constant
\begin{equation}
\tau_0 \equiv \frac{4 \pi \imath}{g^2} + \frac{\vartheta_0}{2 \pi}.
\end{equation}
Indeed, upon evaluating the matrix integral in~\eqref{eq:ZETA} for small $\{M_I\}$, one gets~\cite{Kawai:2004bz}
\begin{equation}\label{eq:z_n1*}
  Z_m^{\N=1^*} = C_{\hN} J_{\hN} \bigg( \frac{2\pi g_m}{\hN g} \bigg)^{\hN^2} \bigg(\frac{2\pi g_m}{\hN M_1 M_2 M_3} \bigg)^{\hN/2}
\big[1 + \dots \big],
\end{equation}
where the dots stand for terms vanishing in the limit $\{M_I\} \to 0$
and $ J_{\hN} = ( \frac{2 \pi \e^{3/2}}{\hN} )^{\frac{\hN^2}{2}}$
is the result of the integration over the angular variables in the $\hat{\Phi}$-integral\footnote{We can diagonalize the matrix $\hat{\Phi}$ so that $\int \du \hat{\Phi} = J_{\hN} \int \prod_i \du \lambda_i \prod_{i<j} (\lambda_i - \lambda_j)^2$.}.

The remarkable fact about eq.~\eqref{eq:z_n1*} is that the $\hN$-leading  contribution to the free-energy, $F_m^{\N=1^*}$ (see eq.~\eqref{eq:FreeE}),
does not depend on $\{M_I\}$. Using eq.~\eqref{eq:FTRI}, one finds
\begin{equation}\label{eq:c_n}
  C_{\hN}=\bigg( \frac{\hN^3 g^2}{(2\pi)^3 \e^{3/2} g_m^2} \bigg)^{\hN^2/2}\exp \left[ -\frac{\pi\imath \tau_0 \hN^2}{ N_c} \right].
\end{equation}
This coefficient, which is a function of $\hN$ and $g_m$, is assumed to be independent of the choice of the potential $W\ped{aux}(\hat{\Phi}_1)$ in eq.~\eqref{eq:mm_sm}.

Exploiting this result, it has been shown explicitly in~\cite{Kawai:2004bz} that choosing $W\ped{aux}(\hat{\Phi}_1)= \um M_1 \hat{\Phi}_1^2$ and in the limit $\{M_I\}\equiv M_0\to\infty$, one can also compute the superpotential of pure $\N=1$ SYM following the Dijkgraaf--Vafa prescription~\cite{Dijkgraaf:2002dh}. The result is
\begin{equation}
W\ped{eff}\api{SYM} = N_c\frac{\partial}{\partial \mathcal{S} }\bigg[ \frac{\mathcal{S}^2}{2} \log \bigg( \frac{\e^{3/2} \Lambda^3}{g^2 \mathcal{S}}\bigg)\bigg]
= N_c \mathcal{S} \bigg[1-\log\bigg(\frac{g^2 \mathcal{S}}{\Lambda^3}\bigg)\bigg]=W_{\rm VY},
\end{equation}
i.e.\ precisely the VY effective potential~\cite{Veneziano:1982ah}.

\subsection{Leigh--Strassler deformations}

The Kawai and co-authors formulation of the Dijkgraaf--Vafa correspondence~\cite{Kawai:2003yf,Kawai:2003dw,Kawai:2004bz} relies on the finiteness of the $\N=4$ theory. Finiteness is not unique to $\N=4$ SYM, as one can concoct various ways (besides adding mass terms~\cite{Kovacs:1999fx,Kovacs:1999rd}) in which it is possible to deform it without losing finiteness.

In 1995 Leigh and Strassler~\cite{Leigh:1995ep} found that it is not uncommon for a supersymmetric theory to display a manifold of fixed points, thus discovering new sets of finite theories. In particular, they were looking for deformations of the $\N=4$ model which could mantain the beta-function and the anomalous dimensions of the chiral superfields vanishing. Amongst them, particular relevance acquired the so-called $\beta$-deformation, which corresponds to a modification of the trilinear coupling in the original $\N=4$ superpotential. Explicitly this modification amounts to the following replacement
\begin{align}\label{eq:N4potential}
 g \tr \Phi_1 [\Phi_2, \Phi_3] \rightarrow h \tr \Phi_1 [\Phi_2, \Phi_3]_\beta, \\
  [X,Y]_\beta \equiv \e^{\imath \beta /2 } XY - \e^{-\imath \beta /2 } YX, \nonumber
\end{align}
where $h$ and $\beta$ are in general complex functions of the coupling constant $g$. The $\N=4$ model is recovered at the point $\beta=0$, $h = g$.

As we said, the deformed model has attracted a lot of interest after its gravity dual was found~\cite{Lunin:2005jy}. In particular, it was shown that the $\beta$-deformed model is finite on a whole submanifold of the parameter space $(h,\beta)$~\cite{Leigh:1995ep,Mauri:2005pa,Rossi:2006mu,Kazakov:2007dy}.

If the scalars have a vanishing vacuum expectation value (vev), $\langle \Phi_I \rangle = 0$, the theory is confining and conformally invariant. In the opposite case, for generic values of the deformation parameters, it is possible to identify regions of the moduli space (branches, in the following) where the gauge group gets spontaneously broken. In order to determine these regions we have to solve the F- and D-flatness conditions which here read
\begin{equation}
  [\Phi_1,\Phi_2]_\beta = [\Phi_2,\Phi_3]_\beta = [\Phi_3,\Phi_1]_\beta = 0
\end{equation}
and
\begin{equation}
  \sum_{I=1}^3 [\Phi_I, \Phi_I^\dagger]=0.
\end{equation}
Unlike the undeformed $\N=4$ case, in which simultaneous diagonalization of the three chiral superfield leads to a solution, now the F-flatness equations are no longer satisfied by arbitrary diagonal matrices, but only by special ones. Setting
\begin{equation}
 \langle \Phi_I \rangle = \diag (\varphi_1^{(I)}, \varphi_2^{(I)}, \dots , \varphi_{N_c}^{(I)}), \qquad I=1,2,3, \label{eq:SETT}
\end{equation}
it follows that at most only one out of the three numbers $\varphi_a^{(1)}, \varphi_a^{(2)}, \varphi_a^{(3)}$ can be non-zero for any $a \in \{ 1, \dots, N_c \}$. In this situation it is customary~\cite{Dorey:2004xm} to introduce the three sets of integers, $\Gamma_I$, which contain the labels of the non-vanishing vev's. They are
\begin{equation}
  \Gamma_I = \big\{ a | \varphi_a^{(I)}\neq 0 \big\}, \qquad a \in \{1,\dots, N_c \}, \quad I=1,2,3,
\end{equation}
and satisfy $\Gamma_I \cap \Gamma_J = \emptyset$ for $I \neq J$. Up to gauge transformation, inequivalent branches are labelled by the triplet $\{n_1,n_2,n_3\}$, where $n_I = \text{ dim } \Gamma_I$. For the sake of illustration we will concentrate in this paper on the particular branches of the type $\{N_c,0,0\}$, where only one of the three chiral superfields (which we take to be $\Phi_1$) develops a vev.

Although at some special point of the moduli space (corresponding to sets of coinciding vev's) some subgroup of the original gauge group can remain unbroken, at a generic point, i.e.\ on what we will be calling the ``Coulomb branch''~\cite{Dorey:2004xm}, the gauge group will be broken spontaneously to $\un{1}^{N_c}$. In this case, the massless spectrum consists of $N_c$ ``photons'', corresponding to the diagonal elements of the gauge field $A_{a}\equiv (V)_{aa}$, and their gluino superpartners $\lambda_{a\alpha} \equiv (\lambda_\alpha)_{aa}$. The two fields combine to form $N_c$ abelian vector supermultiplets of $\N=1$ SUSY. We will denote the corresponding field strength by $w_{a \alpha}$, $a=1,\dots,N_c$. Obviously, there are also $N_c$ massless chiral multiplets, corresponding to the fluctuations around the non-vanishing eigenvalues $\varphi_a^{(1)}$, $a=1,\ldots,N_c$.

What we expect in this situation is that, although no superpotential can be generated, the kinetic term for the massless fields may receive quantum corrections. While not much can be said about the kinetic term of the scalars, as the latter is a D-term, the effective low-energy action for the gauge fields is a \textit{holomorphic} function which must be of the form
\begin{equation}
  W\ped{eff} \propto \sum_{a,b=1}^{N_c} \tau_{ab} w^{\alpha,a} w^{b}_\alpha.
\end{equation}
The complex $N_c \times N_c$ matrix $\tau_{ab}$ encodes the effective gauge couplings and vacuum angles. Restricting to the~$\{N_c,0,0\}$ branch, $\tau_{ab}$ will obviously depend only on the diagonal part of the chiral superfield $\Phi_1$, which according to eq.~\eqref{eq:SETT}, we shall compactly rewrite as $\langle \Phi_1 \rangle$.

At the classical level, $\tau_{ab}$ is proportional to the identity, $\tau_{ab}\api{cl} = \delta_{ab} \tau_0$. Standard non-renormalization theorems guarantee that perturbative quantum effects are limited to one-loop corrections, while non-perturbative instanton-like terms are expected at any order. In other words we will have for $\tau_{ab}(\langle \Phi_1 \rangle) $ an expansion of the type
\begin{equation}\label{eq:tau}
  \tau_{ab}(\langle \Phi_1 \rangle) = \tau_0 \delta_{ab} + \tau\api{1-loop}_{ab} (\langle \Phi_1 \rangle) + \sum_{k=1}^\infty \tau_{ab}^{(k)} (\langle \Phi_1 \rangle) \e^{2 \pi \imath k \tau} \equiv \tau_0 \delta_{ab} + \hat{\tau}_{ab}(\langle \Phi_1 \rangle).
\end{equation}
The one-loop perturbative correction has already been computed in~\cite{Dorey:2004xm,Kuzenko:2005gy}.

\subsubsection{Different formulations of the Leigh--Strassler model}\label{sec:formulations}

In the plain $\N=4$ SYM model we have the freedom to (linearly) redefine the chiral superfields introducing a pair of conjugate fields. Indeed, starting from the usual form of the superpotential $\tr \Phi_1 [\Phi_2 , \Phi_3]$ and defining
\begin{equation}\label{eq:PHIPM23}
  \Phi_{\pm} \equiv \frac{\Phi_2 \pm \imath \Phi_3}{\sqrt{2}},
\end{equation}
the superpotential takes the form $\imath \tr \Phi_1 [\Phi_+ , \Phi_-]$ which up to a $\imath$ factor has the same structure as the one we started from (l.h.s.\ of eq.~\eqref{eq:N4potential}) under the formal replacement $\Phi_{2,3}\to \Phi_{+,-}$. Thus, in the study of the $\N=4$ model it makes no difference whether one considers the $(2,3)$- or the $(\pm)$-formulation of the theory.

However, when $\beta$ is not zero, it is not true anymore that the form of the superpotential is left invariant by the rotation~(\ref{eq:PHIPM23}). In detail, starting from the $(2,3)$-formulation, the $\beta$-deformed superpotential in the r.h.s.\ of eq.~\eqref{eq:N4potential} transforms under the redefinition~\eqref{eq:PHIPM23} as follows
\begin{equation}
  \tr \Phi_1[\Phi_2,\Phi_3]_\beta\to \tr \big(\Phi_1 (\Phi_+^2 - \Phi_-^2) \sin \beta /2 + \imath \Phi_1 [\Phi_+, \Phi_-] \cos \beta/2\big).
\end{equation}
In studying the Leigh--Strassler model a choice must thus be made of which one of the two possible resulting theories one wishes to study. We can either start from the deformed standard superpotential
\begin{equation}\label{eq:SU23}
  W^{(2,3)} = h \Phi_1 [\Phi_2, \Phi_3]_\beta,
\end{equation}
which is what we do in the following, or pick up as superpotential
\begin{equation}\label{eq:SUPM}
  W^{(\pm)} = \imath h \Phi_1 [\Phi_+,\Phi_-]_\beta,
\end{equation}
(obtained by first rotating the fields and then deforming the commutator) which is the choice preferred by other authors~\cite{Dorey:2002pq}. As argued above, the two models are different, or to say it in another way, rotation~\eqref{eq:PHIPM23} and $\beta$-deformation are non-commuting operations. Generically, there is no way of morphing one theory into the other. However, we will show (see appendix~\ref{sec:appA}) that, after adding mass terms for the superfields, in particular limits the two models become equivalent.

\section{The low-energy superpotential of the massive Leigh--Strassler model}\label{sec:computation1}

We now present the results of the MM computation of the effective low-energy superpotential of the massive Leigh--Strassler model in the confining phase of the theory, \textit{i.e.\ }in a situation where the chiral field vev's are zero. In the literature, the model has been exactly solved in the situation  in which the glueball superfield $\mathcal{S}$ is integrated out~\cite{Kostov:1999qx,Berenstein:2002sn,Dorey:2002pq}.

In the Dijkgraaf--Vafa formulation we are required to compute the free-energy of the MM whose action is given by
\begin{equation}\label{eq:SonlyM}
S_{m}(\beta, \{M_I\})=\frac{\hat{N}}{g_m} \tr \left\{ h \hat{\Phi}_1 [\hat{\Phi}_2, \hat{\Phi}_3]_\beta+\sum_{I=1}^3 \frac{M_I}{2} \hat{\Phi}_I^2\right\}.
\end{equation}
While it is clear that, in the limit $\{M_I\} = M_0 \rightarrow \infty$, the trilinear coupling becomes irrelevant, when keeping a finite (but large) value of $M_0$ we expect to recover the correct VY superpotential of pure $\N=1$ Yang-Mills plus a perturbative expansion in terms of inverse powers of $M_0$. This will guarantee that the $M_0 \to \infty$ limit will bring us back to the pure $\N=1$ SYM result.

The computation proceeds as follows. Starting from
\begin{align}\label{eq:3.2}
  Z (\beta, M_0) ={}& \exp\left[-\frac{ \hat{N}^2}{ g_m^2} F_m\right] =  C_{\hat{N}} \int \du \hat{\Phi}_1 \du \hat{\Phi}_2 \du \hat{\Phi}_3 \e^{-S_{m}(\beta, M_0)},
\end{align}
we can immediately integrate over $\hat{\Phi}_2$ and $\hat{\Phi}_3$. After diagonalizing the remaining matrix $\hat{\Phi}_1$, one obtains (see appendix~\ref{app:subleading} for details)
\begin{multline}\label{eq:Zbetafull}
  Z(\beta, M_0) = C_{\hat{N}} J_{\hN} \left[ \frac{ (2 \pi)^2 g^3_m}{\hat{N}^3 M_0^3}\right]^{\frac{\hat{N}^2}{2}} \int \prod_i \du \lambda_i \prod_{i<j} (\lambda_i - \lambda_j)^2\e^{-\sum_i \lambda_i^2 / 2} \\
  \prod_i \big[ 1 + 4 \epsilon \lambda_i^2 \sin^2 \beta /2 \big]^{-\frac{1}{2}} \prod_{i<j} \frac{1}{1 + \epsilon (\lambda_i^2 + \lambda_j^2 -2 \lambda_i \lambda_j \cos \beta)},
\end{multline}
where $\epsilon = \displaystyle{\frac{g_m h^2}{\hat{N} M_0^3}}$.

Terms coming from the product over the $i$ index, $\prod_i(\ldots)$, in the second line of eq.~\eqref{eq:Zbetafull} do not yield leading $\hat{N}^2$-contributions to the MM free-energy, as shown in appendix~\ref{app:subleading}. We are then left with the perturbative expansion of the last product in powers of the small parameter $\epsilon$. The dominant term is
\begin{equation}
  Z_m^{(0)} = C_{\hat{N}} J_{\hN} \left[ \frac{2 \pi g_m}{\hat{N} M_0} \right]^{\frac{3}{2}\hat{N}^2} = \left[ \frac{g^2 g_m}{M_0^3 \e^{3/2}} \right]^{\frac{\hN^2}{2}} \exp \left[- \frac{\pi \imath \tau_0 \hat{N}^2}{N_c}\right],
\end{equation}
from which the ``planar'' free-energy immediately follows
\begin{equation}
  F_m^{(0)} = - \frac{g_m^2}{2} \log \frac{g^2 g_m}{M_0^3 \e^{3/2}} + \frac{\pi \imath \tau_0 g_m^2}{N_c}.
\end{equation}
At this point, application of the Dijkgraaf--Vafa proposal yields for the low-energy effective superpotential the VY result
\begin{align}\label{eq:W23zero}
  W\ped{eff}^{(0)} (\mathcal{S}) \equiv N_c \frac{\partial}{\partial \mathcal{S}}F_m^{(0)} (g_m \to \mathcal{S}) = {} &  N_c \mathcal{S} - N_c \mathcal{S} \log \frac{g^2 \mathcal{S}}{M_0^3} + 2 \pi \imath \tau_0 \mathcal{S} \nonumber \\
 = {} & N_c \mathcal{S} - N_c \mathcal{S} \log \frac{g^2 \mathcal{S}}{\Lambda^3}=W_{\rm VY},
\end{align}
where in rewriting the last equality use was made of the renornalization group running of the complexified gauge coupling constant (eq.~\eqref{eq:Lambda}). The result~\eqref{eq:W23zero} is precisely the expected VY superpotential.

The perturbative expansion of eq.~\eqref{eq:Zbetafull} in terms of the parameter $\epsilon$ gives the higher order corrections to the MM free-energy due to finite mass effects. Explicitely, exploiting the MM identification of $g_m$ with $\mathcal{S}$, one finds
\begin{align}\label{eq:Fbetaunostar}
  F_m - F_m^{(0)} \equiv {} & \Delta F_{m} (\mathcal{S}, \beta, M_0)\nonumber \\
  = {} & -\mathcal{S}^2 \left\{ \left( \frac{h^2 \mathcal{S}}{M_0^3}\right) \right. \nonumber\\
  &  \phantom{-\mathcal{S}^2 \{}  -  \left(\frac{h^2 \mathcal{S}}{M_0^3}\right)^2 \frac{1}{2} (5 + 2 \cos^2 \beta)  \nonumber \\
  & \phantom{-\mathcal{S}^2 \{}  + \left(\frac{h^2 \mathcal{S}}{M_0^3}\right)^3 (11 + 12 \cos^2 \beta)  \nonumber \\
  & \phantom{-\mathcal{S}^2 \{}  - \left(\frac{h^2 \mathcal{S}}{M_0^3}\right)^4 3(21 + 42 \cos^2 \beta + 4 \cos^4 \beta)  \nonumber \\
  & \phantom{-\mathcal{S}^2 \{}  \left. + \left(\frac{h^2 \mathcal{S}}{M_0^3}\right)^5 \frac{4}{5} (527 + 1625 \cos^2 \beta +440 \cos^4 \beta) + \ldots \right\}.
\end{align}
One can check that in the limit $\beta \to 0$, \textit{i.e.} in the pure $\N=1^*$ model, eq.~\eqref{eq:Fbetaunostar} correctly reproduces the known result~\cite{Dijkgraaf:2002pp,Arnone:2007au}
\begin{align}
\Delta F_m (\mathcal{S}, M_0) =  \mathcal{S}^2 \left[ - \big( \frac{g^2 \mathcal{S}}{M_0^3} \big) + \frac{7}{2} \big(\frac{g^2 \mathcal{S}}{M_0^3}\big)^2 - 23  \big(\frac{g^2 \mathcal{S}}{M_0^3}\big)^3 + \ldots \right].
\end{align}
Instead, in the $\beta \neq 0$ case the contribution to the effective superpotential is obtained by adding eq.~\eqref{eq:W23zero} to the contribution coming from eq.~\eqref{eq:Fbetaunostar}, and reads (with the superscript $^{(2,3)}$ we specify that we are dealing with the (2,3)-formulation (eq.~(\ref{eq:SU23})) of $\beta$d SYM)
\begin{multline}\label{eq:W23pert}
  W\ped{eff}^{(2,3)}(\mathcal{S},\beta,M_0) = N_c \mathcal{S} - \mathcal{S} \log \Big[\frac{g^2 \mathcal{S}}{\Lambda^3}\Big]^{N_c} - 3 N_c \frac{h^2 \mathcal{S}^2}{M_0^3} + 2 N_c  \frac{h^4 \mathcal{S}^3}{M_0^6} (5 + 2 \cos^2 \beta) \\
  - 5 \frac{h^6 \mathcal{S}^4}{M_0^9} (11 + 12 \cos^2 \beta) + \dots
\end{multline}
If we eliminate $\mathcal{S}$ by the condition which extremizes the superpotential,
\begin{equation}
  \frac{\partial W^{(2,3)}\ped{eff}}{\partial \mathcal{S}} = 0,
\end{equation}
yielding $W^{(2,3)}\ped{on-shell}$, we can compare our result with the on-shell formula obtained in~\cite{Dorey:2002pq}, which at \textit{all orders} in the ($\pm$)-formulation takes the form
\begin{equation}\label{eq:ALLOR}
  W\ped{on-shell}^{(\pm)}(\beta, M_0) = \frac{N_c M_0^3}{4 h^2 \sin^2 \beta/2}- \frac{N_c M_0^3 \cos \beta/2}{4 h^2 \sin^3 \beta/2}  \frac{\theta_1 (\beta/2 | \tau\ped{R}/N_c)}{\theta'_1(\beta/2| \tau\ped{R}/N_c)},
\end{equation}
where $\theta_1 (z | \tau)$ is one of the standard Jacobi elliptic functions~\cite{whittakerwatson:1927} and
\begin{equation}
  \tau\ped{R} = \tau_0 - \frac{\imath N_c}{\pi} \ln h. \label{TR}
\end{equation}
As signaled by the superscript ($\pm$) attached to the superpotential and as already noted in sec.~\ref{sec:formulations}, the authors of this paper have worked in the $(\pm)$-formulation of the (undeformed) $\N=4$ SYM model. Then, in order to be able to make contact with their result, we must repeat our previous calculation starting from the expression~\eqref{eq:SUPM} of the superpotential.

If we do so, the result for the free-energy differs from eq.~(\ref{eq:Fbetaunostar}) and reads
\begin{equation}
  F_m^{(\pm)} = \frac{\pi \imath \tau_0}{N_c}\mathcal{S}^2 - \frac{\mathcal{S}^2}{2} \log \left[ \frac{g^2 \mathcal{S}}{M_0^3 \e^{3/2} }\right] - \frac{h^2 \mathcal{S}^3}{M_0^3}(1 - 2\cos \beta)+\dots, \label{FRE}
\end{equation}
from which we obtain for the effective superpotential in the ($\pm$)-formulation the $1/M_0$-expansion
\begin{equation}\label{eq:Wpmnostro}
  W^{(\pm)}\ped{eff}(\mathcal{S},\beta,M_0) = N_c \mathcal{S} - \mathcal{S} \log \Big[\frac{g^2 \mathcal{S}}{\Lambda^3}\Big]^{N_c} - 3 N_c \frac{h^2 \mathcal{S}^2}{M_0^3} (1 - 2\cos \beta) + \dots.
\end{equation}
From the discussion above, it should be no surprise that this result is different from the one we obtained considering the $(2,3)$-form of the superpotential, given in eq.~\eqref{eq:W23pert}.

Now we can finally draw comparisons between the MM computation of the superpotential and the result of~\cite{Dorey:2002pq}. In fact, starting from eq.~\eqref{eq:Wpmnostro}, we can eliminate $\mathcal{S}$ by extremizing the superpotential~(\ref{eq:Wpmnostro}), finding for the on-shell effective superpotential the expansion
\begin{equation}\label{eq:Wpmonshell}
  W^{(\pm)}\ped{eff} (\beta,M_0)= \frac{N_c M_0^3}{g^2} q^2 + 3 \frac{N_c h^2 M_0^3}{g^4} q^4(2 \cos \beta -1)+ \mathcal{O}(q^6), \quad q \equiv \e^{\pi \imath \tau_0},
\end{equation}
which coincides with the first few terms in eq.~\eqref{eq:ALLOR}. Letting $M_0 \to 0$ produces a vanishing contribution. This is to be expected since in this limit we should recover pure $\N=4$ Yang-Mills whose only holomorphic contribution to the superpotential is the bare kinematical term of the action.

We have already remarked how, in the $\beta \to 0$ limit, the $(2,3)$- and the $(\pm)$-formulations coincide (and are equivalent to the $\N=1^*$ model). Indeed, if we integrate out $\mathcal{S}$ from the effective superpotential~\eqref{eq:Fbetaunostar} and then send $\beta$ to zero we have the same result we get putting $\beta \to 0$ in eq.~\eqref{eq:Wpmonshell}, namely the Eisenstein series
\begin{align}
  W\ped{eff}^{(\pm)}(\beta = 0) = {} & N_c \frac{M_0^3}{g^2} \big[ q^2 + 3 q^4 + 4 q^6 + 7 q^8 + 6 q^{10} +\mathcal{O}(q^{12}) \big] \\
  = {} &  W\ped{eff}^{(2,3)} (\beta=0). \nonumber
\end{align}

Another interesting result (proved in appendix~\ref{sec:appA}) is that, also for non-vanishing $\beta$, in the MM formalism the massless limit of the superpotential of the massive Leigh--Strassler model does not depend on whether the ($\pm$)- or the ($2,3$)-formulation is used.

In conclusion, the web of relations between the two formulations in the various limits considered can be schematically represented by the following flow-chart
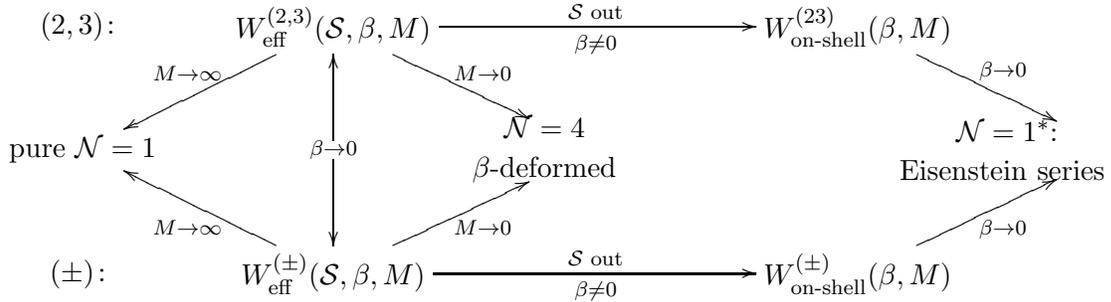
\begin{displaymath}
  \xymatrix{(2,3) \colon & \ar[dl]_{M \to \infty} W\ped{eff}^{(2,3)} (\mathcal{S},\beta,M) \ar@{<->}[dd]|{\beta \to 0} \ar[rr]^{\mathcal{S} \text{ out}}_{\beta \neq 0} \ar[dr]^{M \to 0} & & W\ped{on-shell}^{(23)} (\beta, M) \ar[dr]^{\beta \to 0}  & \\
  \text{pure } \mathcal{N}=1 & &   \parbox{6em}{\centering \hspace{-4em} $\N = 4$ \\ \hspace{-4em} $\beta$-deformed}&  & \hspace{-2em} \parbox{8em}{\centering \hspace{-6em} $\N = 1^*$:\\ \hspace{-6em}Eisenstein series} \\
 (\pm) \colon & \ar[ul]^{M \to \infty} W\ped{eff}^{(\pm)} (\mathcal{S},\beta,M) \ar[rr]^{\mathcal{S} \text{ out}}_{\beta \neq 0} \ar[ur]_{M \to0} &  & W\ped{on-shell}^{(\pm)} (\beta, M) \ar[ur]_{\beta \to 0}  & }
\end{displaymath}

\section{Spontaneous symmetry breaking in the $\N=4$ Leigh--Strassler model}\label{sec:computation2}

In this section we investigate the behaviour of the massive Leigh--Strassler deformation of the $\N=4$ SYM when the gauge group $\un{N_c}$ is spontaneously broken. By varying the parameters $(\beta, M_0)$ different models are encountered. In particular, when $M_0$ goes to zero, we expect to recover the $\beta$d SYM theory, while for $\beta \to 0$ we should go back to the $\N=1^*$ model. Moreover, the limit in which both $\beta$ and $M_0$ vanish must reproduce pure $\N=4$ SYM, whose holomorphic contribution to the effective superpotential is, as we repeatedly said, just its tree-level kinematical term. As such, it can be used as a sort of boundary condition in parameter space useful to constrain the $(\beta, M_0)$ dependence of the holomorphic superpotential in more general situations.

If one wishes to induce a spontaneous symmetry-breaking of the form
\begin{equation}\label{eq:SSB}
\un{N_c} \mapsto \prod_{i=1}^n \un{N_i}, \qquad  \sum_{i=1}^n N_i = N_c,
\end{equation}
one has to include in the MM action an auxiliary potential $W\ped{aux}(\hat\Phi_1)$ of degree $n+1$. In the $\hN \to \infty$ limit the dominant configuration in the (zero-dimensional) functional integral will be the one in which the matrix eigenvalues are distributed among the $n$ extrema of the $W\ped{aux}$ potential, with $\hN_i$ eigenvalues located around the $i$-th extremum. The integers $\hN_i$  satisfy the constraint $\hN = \hN_1 + \dots + \hN_n$ and will be also taken to grow infinitely large, proportionally to $\hN$ as $\hN \to \infty$.

For every gauge group factor $\un{N_i}$ we have the choice of defining a glueball superfield for the full group or for the $\sun{N_i}$ subgroup only, namely
\begin{equation}
\mathcal{S}_i = - \frac{1}{32 \pi^2} \Tr\ped{$\un{N_i}$} \mathcal{W}_{i}^{\alpha} \mathcal{W}_{i,\alpha}, \quad \hat{\mathcal{S}}_i = - \frac{1}{32 \pi^2} \Tr\ped{$\sun{N_i}$} \mathcal{W}_{i}^{\alpha} \mathcal{W}_{i,\alpha},
\end{equation}
where $\mathcal{W}_{i}^{\alpha}$ is the supersymmetric gauge field strength. Obviously, it is $\hat{\mathcal{S}}_i$ that describes confinement and gaugino condensation, the abelian $\un{1}$ degrees of freedom being IR-free. However, the form of the Dijkgraaf--Vafa correspondence is particularly simple only if expressed in terms of the full glueball superfields, $\mathcal{S}_i$, corresponding to the $\un{N_i}$ gauge groups. In the MM setup, the $\mathcal{S}_i$'s are the objects which are identified by the correspondence
\begin{equation}
  \mathcal{S}_i \Leftrightarrow g_m \frac{\hN_i}{\hN},
\end{equation}
once the large-$\hN$ limit is attained. Thus, for a generic symmetry-breaking pattern, the Dijkgraaf--Vafa recipe gives, for the $\mathcal{S}$-dependent part of the effective superpotential, the expression
\begin{equation}
  W\ped{eff} (\mathcal{S}) = \sum_{i=1}^n N_i \frac{\partial F_m (\mathcal{S})}{\partial \mathcal{S}_i},
\end{equation}
while, apart from the tree-level term $\tau_0\delta_{ij}$, the ``coupling constant matrix'' of the massless $\un{1}$'s degrees of freedom is given by (see eq.~\eqref{eq:tau})
\begin{equation}\label{eq:taugeneral}
  \hat{\tau}_{ij} = \frac{\partial^2 F_m(\mathcal{S})}{\partial \mathcal{S}_i \partial \mathcal{S}_j} - \delta_{ij} \frac{1}{N_i} \sum_{k=1}^n N_k \frac{\partial^2 F_m (\mathcal{S})}{\partial S_i \partial \mathcal{S}_k}.
\end{equation}
To obtain the final interesting expression of the coupling constant matrix the last necessary step is to extremize the effective superpotential, \textit{i.e.} solve the equations
\begin{equation}
  \frac{\partial W\ped{eff}}{\partial \mathcal{S}} = 0.
\end{equation}
The latter is a system of $n$ equations in the $n$ unknowns $\mathcal{S}_i$ which, once expressed in terms of the parameters of the model, will have to be put back into eq.~\eqref{eq:taugeneral}.

In the following, to simplify formulae, we will concentrate on the branch identified by the triplet $\{N_c,0,0\}$, \textit{i.e.} to the case in which the gauge group is $\un{N_c}$ broken down to $\un{N_1} \times \un{N_2}$ with $N_1+N_2=N_c$, by a non-vanishing $\langle \Phi_1\rangle$. This situation corresponds to the following particular form of the MM partition function
\begin{equation}\label{eq:Zbeta}
  Z_m =\exp\left[-\frac{ \hat{N}^2}{ g_m^2} F_m\right] = C_{\hat{N}} \int \du \hat{\Phi}_1 \du \hat{\Phi}_2 \du \hat{\Phi}_3  \e^{-S_m (\hat{\Phi}_I;\beta,M_0)},
\end{equation}\begin{equation}
  S_m(\hat{\Phi}_I;\beta,M_0) = \frac{\hat{N}}{g_m} \tr \left\{  h \hat{\Phi}_1 [\hat{\Phi}_2, \hat{\Phi}_3]_\beta + \frac{M_0}{2} (\hat{\Phi}_2^2 + \hat{\Phi}_3^2) + W\ped{aux}(\hat{\Phi}_1)\right\}.
\end{equation}
with
\begin{equation}
W\ped{aux}( \hat{\Phi}_1) = \gamma \left[ \hat{\Phi}_1^3 - \um (\varphi_1 + \varphi_2) \hat{\Phi}_1^2 + \varphi_1 \varphi_2 \hat{\Phi}_1 \right],
\end{equation}
where $\varphi_{1,2}$ ($\varphi_{1}\neq \varphi_{2}$) are the eigenvalues of $\hat{\Phi}_1$ (see eq.~\eqref{eq:SETT}) identifying the point on the branch we are interested in. The choice of a cubic potential (which has two stationary points) allows us to describe the $\un{N_c} \mapsto \un{N_1} \times \un{N_2}$ symmetry breaking pattern. Since $W\ped{aux}( \hat{\Phi}_1)$ has only the r\^ole of inducing the spontaneous breaking of the $\un{N_c}$ gauge symmetry, we expect physical quantities be independent of the magnitude of the breaking potential. In other words the final answer should not depend on $\gamma$.

After integrating the quadratic $\hat{\Phi}_2$ and $\hat{\Phi}_3$ dependence, diagonalization of the remaining $\hat{\Phi}_1$ matrix leads to the formula
\begin{multline}\label{eq:Z123}
  Z_m = \e^{-\pi \imath \tau_0 \hN^2/N_c} \left[\frac{g^2}{h^2} \right]^{\frac{\hN^2}{2}} \int \prod_{i=1}^{\hN} \du \lambda_i \exp\left[-\frac{ \hN}{g_m} \sum_i W\ped{aux}(\lambda_i)\right]  \\
  \prod_i \left[ \frac{M_0^2}{h^2} + 4 \lambda_i^2 \sin^2 \beta/2\right]^{-1} \prod_{i<j} \frac{(\lambda_i - \lambda_j)^2}{\frac{M_0^2}{h^2} + (\lambda_i^2 + \lambda_j^2 - 2 \lambda_i \lambda_j \cos \beta)} ,
\end{multline}
where we have inserted the known expression for $C_{\hat{N}}$ given in eq.~\eqref{eq:c_n}.

At this point, the standard procedure consists in expanding each eigenvalue, $\lambda_i$, around either $\varphi_{1}$ or  $\varphi_{2}$ which are the extrema of the potential. We then write
\begin{equation}
  \lambda_i = \left\{
    \begin{array}{ll}
      \varphi_1 + p_i & i = 1,\dotsc, \hN_1 \\
      \varphi_2 + p_i & i = \hN_1 + 1,\dotsc, \hN_1 + \hN_2
    \end{array}\right.
  \qquad \hN_1 + \hN_2 = \hN
\end{equation}
For convenience, in the following we will set $k = i - \hN_1$ and $p_i \equiv q_k$, whenever $i>\hN_1$. Then, the symmetry breaking potential becomes
\begin{equation}
  -\frac{\hN}{g_m} W\ped{aux}(\lambda_i) \to \left\{
    \begin{array}{lr}
      \displaystyle{- \um \frac{\hN}{g_m} \gamma (\varphi_1 - \varphi_2) p_i^2
- \frac{1}{3} \frac{\hN}{g_m} \gamma p_i^3} & \quad i=1,\dotsc, \hN_1 \\
      \displaystyle{- \um \frac{\hN}{g_m} \gamma (\varphi_2 - \varphi_1) q_k^2
- \frac{1}{3} \frac{\hN}{g_m} \gamma q_k^3} & \quad k=1,\dotsc, \hN_2
    \end{array}
  \right.  ,
\end{equation}
where $\hN_1$ and $\hN_2$ are the numbers of eigenvalues chosen to lie around $\varphi_1$ and $\varphi_2$, respectively. Defining
\begin{equation}\label{eq:XI}
  \xi^{-2} \equiv \gamma (\varphi_1 - \varphi_2) \frac{\hN}{g_m}
\end{equation}
and rescaling the $p_i$ and $q_k$ variables in order to recover the standard gaussian weight by putting
\begin{equation}
  \begin{array}{l}
    \displaystyle{p_i^2 \rightarrow p_i'^2 = \frac{p_i^2}{\xi^2} \quad \Rightarrow \quad \gamma \frac{\hN}{g_m} (\varphi_1 - \varphi_2 ) p_i^2 \rightarrow p_i^2}, \\
    \displaystyle{q_k^2 \rightarrow q_k'^2 = - \frac{q_k^2}{\xi^2} \quad \Rightarrow \quad \gamma \frac{\hN}{g_m} (\varphi_2 - \varphi_1 ) q_k^2 \rightarrow - q_k^2},
  \end{array}
\end{equation}
one gets
\begin{align}\label{eq:Zprimonoexp}
  &Z_m= \e^{-\pi \imath \tau_0 \hN^2 /N_c} \left[\frac{g^2}{h^2}\right]^{\frac{\hN^2}{2}} \big(\xi\big)^{\hN_1^2} \big(- \xi\big)^{\hN_2^2} \int \prod_{i=1}^{\hN_1} \du p_i \prod_{k=1}^{ \hN_2} \du q_k \prod_{i<j}^{\hN_1} (p_i -p_j)^2 \prod_{k<l} (q_k - q_l)^2 \nonumber\\
  &\prod_{i,k}^{\hN_1,\hN_2} \frac{\big[ \xi p_i + \imath \xi q_k + \varphi_1 - \varphi_2 \big]^2}{\big[\frac{M_0^2}{h^2} + [(\xi p_i + \varphi_1) \e^{\imath \beta/2} - (\varphi_2 -\imath\xi q_k)\e^{-\imath \beta/2}][(\xi p_i + \varphi_1) \e^{-\imath \beta/2} - (\varphi_2 -\imath\xi q_k)\e^{\imath \beta/2}]\big]} \nonumber\\
  &\prod_{i<j}^{\hN_1} \Big[ \frac{M_0^2}{h^2} + [(\xi p_i + \varphi_1)\e^{\imath \beta/2} - (\xi p_j + \varphi_1)\e^{-\imath \beta/2}][(\xi p_i + \varphi_1)\e^{-\imath \beta/2} - (\xi p_j + \varphi_1)\e^{\imath \beta/2}] \Big]^{-1} \nonumber \\
  &\prod_{k<l}^{\hN_2} \Big[ \frac{M_0^2}{h^2} + [(\varphi_2 -\imath \xi q_k ) \e^{\imath \beta/2} - (\varphi_2 - \imath \xi q_l)\e^{-\imath \beta/2}][(\varphi_2 - \imath \xi q_k)\e^{-\imath \beta/2} - (\varphi_2 - \imath \xi q_l) \e^{\imath \beta/2}] \Big]^{-1} \nonumber \\
  & \exp \Big\{-\um \Big[\sum_i^{\hN_1} {p}_i^2 + \sum_k^{\hN_2} {q}_k^2\Big] - \frac{\xi}{3 (\varphi_1 - \varphi_2)}\Big[\sum_i^{\hN_1} {p}_i^3 + \imath \sum_k^{\hN_2} q_k^3\Big] \Big\}.
\end{align}

Since in the resulting action the $q_k$'s were displaying an effective negative mass squared, we suitably deformed the contour to make the integral convergent, by a sort of Wick rotation, as is usual in the Dijkgraaf--Vafa approach. Besides, we do not include the contribution coming from the $\prod_i$ product in (the first term of the second line of) eq.~\eqref{eq:Z123}, as it does not lead to relevant terms in the large-$\hN$ limit, as shown in appendix~\ref{app:subleading}. The expression~\eqref{eq:Zprimonoexp} will be the starting point of the analysis presented in the following sections.

\subsection{Matrix model perturbative expansion}

The lowest order term is produced by a straight Gaussian integration in eq.~\eqref{eq:Zprimonoexp}, yielding
\begin{align}
   & Z_m^{(0)} = \e^{-\pi \imath \tau_0 \hN^2 /N_c} \left[ \frac{M_0^2}{h^2} - \varphi_1^2(\e^{\imath \beta /2 } - \e^{- \imath \beta /2})^2 \right]^{-\frac{\hN_1^2}{2}} \left[ \frac{M_0^2}{h^2} - \varphi_2^2 (\e^{\imath \beta /2 } - \e^{- \imath \beta /2})^2 \right]^{-\frac{\hN_2^2}{2}} \nonumber \\
   &\left[ \frac{g^2( \varphi_1 - \varphi_2)^2}{M_0^2 + h^2 (\e^{\imath \beta/2} \varphi_1 - \e^{-\imath \beta/2} \varphi_2)(\e^{-\imath \beta/2} \varphi_1 - \e^{\imath \beta/2} \varphi_2)} \right]^{\hN_1 \hN_2}  \left( \frac{g^2}{h^2}\right)^{\frac{\hN_1^2 +\hN_2^2}{2}} (\xi)^{\hN_1^2} (-\xi)^{\hN_2^2} \nonumber \\
   &\int \prod_{i=1}^{\hN_1} \du p_i \prod_{k=1}^{\hN_2} \du q_k \prod_{i<j}^{\hN_1} (p_i - p_j)^2 \prod_{k<l}^{\hN_2} (q_k - q_l)^2 \exp \Big\{-\um \Big[\sum_i^{\hN_1} {p}_i^2 + \sum_k^{\hN_2} {q}_k^2\Big]\Big\} \nonumber\\
   & = \e^{-\pi \imath \tau_0 \hN^2/N_c} \left[ \frac{M_0^2}{h^2}  - \varphi_1^2(\e^{\imath \beta /2 } - \e^{- \imath \beta /2})^2 \right]^{- \frac{\hN_1^2}{2}} \left[ \frac{M_0^2}{h^2} - \varphi_2^2(\e^{\imath \beta /2 } - \e^{- \imath \beta /2})^2 \right]^{- \frac{\hN_2^2}{2}} (\xi)^{\hN_1^2} (-\xi)^{\hN_2^2} \nonumber \\
   & \left[ \frac{g^2( \varphi_1 - \varphi_2)^2}{M_0^2 + h^2(\e^{\imath \beta/2} \varphi_1 - \e^{-\imath \beta/2} \varphi_2)(\e^{-\imath \beta/2} \varphi_1 - \e^{\imath \beta/2} \varphi_2)} \right]^{\hN_1 \hN_2} \Bigg( \frac{g^2 \hN_1}{h^2 \e^{3/2}}\Bigg)^{\frac{\hN_1^2}{2}}  \Bigg(\frac{g^2 \hN_2}{h^2 \e^{3/2}}\Bigg)^{\frac{\hN_2^2}{2}}, \nonumber
\end{align}
implying for the leading (tree-level + one-loop) contribution to the free-energy the formula
\begin{align}\label{eq:Fzeroth}
  F_m^{(0)} = \frac{\pi \imath \tau_0}{N_c} (\mathcal{S}_1 + \mathcal{S}_2)^2 - \sum_{i=1}^2 \frac{\mathcal{S}_i^2}{2} \log \left[ \frac{g^2 \mathcal{S}_i}{h^2 K_i (\varphi_1 - \varphi_2) \e^{3/2}} \right] - \mathcal{S}_1 \mathcal{S}_2 \log \frac{g^2 \Delta^2(0)}{h^2 \Delta(\beta)\Delta(-\beta) + M_0^2}.
\end{align}
In eq.~(\ref{eq:Fzeroth}) we have introduced the definitions
\begin{equation}
K_i \equiv (-)^i \gamma \left[ \frac{M_0^2}{h^2} - \varphi_i^2 (\e^{\imath \beta /2} - \e^{-\imath \beta /2})^2 \right]  \quad \text{and} \quad \Delta(x) \equiv \e^{\imath x /2} \varphi_1 - \e^{-\imath x /2} \varphi_2.
\end{equation}
Following the Dijkgraaf--Vafa prescription we can derive from eq.~\eqref{eq:Fzeroth} the coupling constant matrix for the two massless abelian fields corresponding to the two U(1) subgroups of the unbroken $\un{N_1} \times \un{N_2}$ gauge symmetry. From the formula (see eq.~(\ref{eq:taugeneral}))~\cite{Cachazo:2002ry}
\begin{align}
\hspace{-.2cm}\hat{\tau} = \tau \begin{pmatrix} -\frac{N_2}{N_1} & 1 \\ 1 & -\frac{N_2}{N_1} \end{pmatrix},
\quad\tau = \frac{\partial^2 F_m(\mathcal{S}_i)}{\partial \mathcal{S}_1 \partial \mathcal{S}_2}=\tau\api{1-loop} + \tau^{(1)} + \tau^{(2)} + \dots, \label{eq:tauhat2}
\end{align}
we get from eq.~(\ref{eq:Fzeroth}) the 1-loop expression
\begin{equation}
  \tau\api{1-loop} = - \log \left[ \frac{g^2( \varphi_1 - \varphi_2)^2}{M_0^2 + h^2(\e^{\imath \beta/2} \varphi_1 - \e^{-\imath \beta/2} \varphi_2)(\e^{-\imath \beta/2} \varphi_1 - \e^{\imath \beta/2} \varphi_2)} \right],
\end{equation}
which in the $M_0\to 0$ limit correctly reproduces the known results of~\cite{Dorey:2004xm,Kuzenko:2005gy}.

The constraint $\sum_i N_i \hat{\tau}_{ij} = 0$, which is automatically satisfied by the definition~\eqref{eq:taugeneral}, is nothing but the condition ensuring the complete decoupling of the overall diagonal $\un{1}$ factor contained in the original $\un{N_c}$ gauge group.

\subsection{Next order(s)}

We can carry on our perturbative treatment expanding eq.~\eqref{eq:Zprimonoexp} to higher orders in the small parameter $\xi$ (eq.~\eqref{eq:XI}). One can get in this way the contribution to the coupling constants matrix up to $n$ instantons.

Stopping at second order in the MM formulation (\textit{i.e.} to order $\xi^4$), the complete expression of $\tau$, upon elimination of $\mathcal{S}_1$ and $\mathcal{S}_2$ via the standard effective superpotential extremization, is given by the quite complicated formula
\begin{align}
  &\tau\api{1-loop} + \tau^{(1)} + \tau^{(2)}= -\log \frac{g^2 (\varphi_1 - \varphi_2)^2}{h^2 (\e^{\imath \beta /2 }\varphi_1 - \e^{- \imath \beta /2} \varphi_2)(\e^{-\imath \beta /2 }\varphi_1 - \e^{\imath \beta /2} \varphi_2)}  \nonumber \\
  &\hspace{4em} + \frac{8 h^4 A^2 \sin^2\beta/2}{g^4 \Delta^4(0)\Delta(\beta) \Delta(-\beta)} \big[ (-3 + 4\cos \beta -\cos 2\beta )\varphi_1^6 + 2 (-5 + 6 \cos \beta - \cos 2 \beta) \varphi_1^5 \varphi_2 \nonumber \\
  &\hspace{4em} +  (-13 + 16 \cos \beta - 3 \cos 2\beta) \varphi_1^4 \varphi_2^2 + 8 (-1 + 2\cos \beta - \cos 2\beta) \varphi_1^3 \varphi_2^3  \nonumber \\
  &\hspace{4em} +  (-13 + 16 \cos \beta - 3 \cos 2\beta) \varphi_1^2 \varphi_2^4 + 2 (-5 + 6 \cos \beta - \cos 2 \beta)\varphi_1 \varphi_2^5 \nonumber \\
  &\hspace{4em} +  (-3 + 4\cos \beta -\cos 2\beta )\varphi_2^6 \big] \nonumber \\
  &\hspace{4em} +\frac{16 h^8 A^4 \sin^4 \beta/2}{g^8 \Delta^8(0)\Delta^2(\beta) \Delta^2(-\beta)} \big[(-139 + 166 \cos \beta - 6 \cos 2 \beta - 26 \cos 3 \beta + 5 \cos 4 \beta ) \varphi_1^{12}\nonumber \\
  &\hspace{4em}  +2 (-258+368 \cos \beta -125 \cos 2 \beta + 30\cos 3 \beta + 3 \cos 4  \beta ) \varphi_1^{11} \varphi_2 \nonumber\\
  &\hspace{4em} - 2(446-767 \cos \beta +464  \cos 2  \beta-89  \cos 3  \beta + 36  \cos 4 \beta ) \varphi_1^{10} \varphi_2^2\nonumber \\
  &\hspace{4em}  +2 (-606+1116 \cos\beta - 709 \cos 2 \beta +250 \cos 3 \beta +39  \cos 4  \beta ) \varphi_1^9 \varphi_2^3\nonumber \\
  &\hspace{4em} +(-2513+3650 \cos \beta -870 \cos 2 \beta +578  \cos 3 \beta - 125 \cos 4 \beta ) \varphi_1^8 \varphi_2^4\nonumber 
\end{align}
\begin{align}\label{eq:T12}
  & \hspace{-1em} -4(876-1220 \cos \beta + 871 \cos 2 \beta +6 \cos 3 \beta +97 \cos 4 \beta ) \varphi_1^7 \varphi_2^5\nonumber \\
  & \hspace{-1em} +2 (-827 + 1745 \cos \beta -694 \cos 2 \beta +559 \cos 3 \beta + 99 \cos 4 \beta) \varphi_1^6 \varphi_2^6\nonumber \\
  & \hspace{-1em} + (\varphi_1 \leftrightarrow \varphi_2) \Big]
\end{align}
As expected, eq.~(\ref{eq:T12}) is organized as a power series expansion in the instanton action,
\begin{equation}
  A^{2} \equiv \exp [2 \pi \imath \tau_0] \propto \exp [- 8 \pi^2 / g^2 ].
\end{equation}
Remebering that the $\N=4$ SYM theory is recovered in the $h^2 \rightarrow g^2$ and $\beta \to 0$ limit, we can easily check that
\begin{equation}
  \lim_{\beta \to 0} \tau = 0,
\end{equation}
leaving only the overall diagonal contribution, $\tau_0$ (see eq.~\eqref{eq:tau}). We stress that, as expected, the matrix $\tau_{ij}$ does not depend on the strength of the symmetry-breaking potential, $\gamma$. This is a quite satisfactory result which confirms our interpretation of the physics described by the formalism as a spontaneous symmetry breaking phenomenon.

\section{Conclusions}\label{sec:conclusions}

In this paper, we have explored some aspects of the deep connection between matrix models and $\N=1$ supersymmetric gauge field theories.

We have started considering the massive Leigh--Strassler model, in two of its most commonly studied formulations, and computed its low-energy effective superpotential as a function of the glueball superfield, for arbitrary (real) values of mass and $\beta$ parameters, exploiting the Dijkgraaf--Vafa conjecture.

We successfully made contact with known results when our formulae are restricted to special points of the parameter space, thus providing new evidence for the existence of a useful correspondence between supersymmetric gauge theories and matrix models. Along the way in this study we have been able to identify the complicated web of relations between the different formulations of the Leigh--Strassler model in various limiting situations.

Then we turned to the study of the spontaneously broken phase of the Leigh--Strassler model, introducing an auxiliary potential term to give a non-vanishing vev to one of the chiral fields. The MM formalism allowed a detailed and general study of the phenomenon of spontaneous breaking of the gauge symmetry. For the sake of simplicity, we displayed the expression of the coupling constant matrix governing the dynamics of the left-over massless degrees of freedom in the simple case of the $\un{N_c}\mapsto \un{N_1} \times \un{N_2}$ symmetry-breaking pattern. We went up to two instantons in the calculation (but there is no problem of principles to go higher in the instanton number), finding agreement with the leading order (tree-level + 1-loop) expression known in the
literature~\cite{Dorey:2004xm,Kuzenko:2005gy}.
Reassuringly our final formulae do not depend on the ``strength'' of the auxiliary potential employed to induce the phenomenon of gauge symmetry breaking, confirming in this way Dijkgraaf and Vafa expectations.

A challenging open problem which we leave for a future investigation is the extension the MM approach to more general deformations of $\N=4$ SYM, such as $\Tr \hat\Phi_1^3$ and the like.

\acknowledgments

Is is a pleasure to thank M.~Testa for numerous enlightening  discussions and especially S.~Arnone and G.~Di~Segni who contributed to the early stages of this work.
Part of this work has been done while M.S. was a guest at RIKEN (Wako-Shi, Tokyo, Japan). He would like to thank T.~Tada for his warm hospitality and M.~Ninomiya for advice. Financial support from INFN is gratefully acknowledged. The work of G.C.R. is supported in part by the Italian MIUR-PRIN contract 2006-025843. The work of Ya.S.S. is supported in part by the Italian MIUR-PRIN contract 2007-024045.

\appendix

\section[Large $\hN$ leading and subleading terms]{Large-$\boldsymbol{\hN}$ leading and subleading terms}\label{app:subleading}

In this appendix, we would like to clarify in which sense, in eq.~\eqref{eq:Zbetafull}, in the large-$\hat{N}$ limit we are allowed to forget the term
\begin{equation}
\prod_i \Big[ 1 + 4 \epsilon \lambda_i^2 \sin^2 \beta/2 \Big]^{-\frac{1}{2}}.
\end{equation}
Loosely speaking, this is related to the fact that this factor corresponds to a product of only $\hN$ terms, while it is the contribution from the remaining (much more numerous) $\hN^2 - \hN$ ones that matters. The two terms come from separating the originally unconstrained product $\prod_{i,j}$ into an (irrelevant) $i=j$ piece and the set of $i \neq j$ terms.

The starting point of the argument is the MM formulation of the massive Leigh--Strassler model in its confining phase, whose action in the (2,3)-formulation is given in eq.~\eqref{eq:SonlyM}. The discussion below is limited to the case where $W\ped{aux}$ is quadratic, but it could be extended to a general potential of the form
\begin{equation}
  W(x) = \sum_{k=2}^n c_k x^k.
\end{equation}
Starting from eq.~\eqref{eq:3.2}, we integrate out two of the three matrices and diagonalize the remaining one. After rescaling the eigenvalues by putting
\begin{equation}
  \lambda_i \rightarrow \sqrt{\frac{g_m}{\hN M_0}} \lambda_i,
\end{equation}
we get
\begin{align}\label{eq:Z23appendixC}
  Z(\beta, M_0) = {}& C_{\hN} J_{\hN} \left[ \frac{(2 \pi)^2 g_m^3}{\hN^3 M_0^3} \right]^{\frac{\hN^2}{2}} \int \prod_i \du \lambda_i \prod_{i<j} (\lambda_i - \lambda_j)^2 \e^{- \um \sum_i \lambda_i^2} \nonumber \\
  & \hspace{1em} \prod_i \big[ 1 + 4 \epsilon \lambda_i^2 \sin^2 \beta/2 \big]^{-\um} \prod_{i \neq j} \Big[ 1 + \epsilon (\lambda_i^2 + \lambda_j^2 - 2 \lambda_i \lambda_j \cos \beta) \Big]^{-\um} \nonumber \\
  \equiv & {}  C_{\hN} J_{\hN} \left[ \frac{(2 \pi)^2 g_m^3}{\hN^3 M_0^3} \right]^{\frac{\hN^2}{2}} \big\langle \big\langle \prod_i \cdot \prod_{i\neq j} \big\rangle \big\rangle, \qquad \epsilon \equiv \frac{g_m h^2}{\hN M_0^3},
\end{align}
where the symbols $\prod_i$ and $\prod_{i\neq j}$ in the last line stand for the two factors in the second line of eq.~\eqref{eq:Z23appendixC} and to shorten future formulae we have introduced the ``mean value''-like notation
\begin{equation}\label{eq:TRF}
\big\langle \big\langle \bigstar \big\rangle \big\rangle \equiv \int \prod_i \du \lambda_i \prod_{i<j} (\lambda_i - \lambda_j)^2 \e^{-\um \sum_i \lambda_i^2} \bigstar.
\end{equation}
Our purpose is to compute the logarithm of $Z$ (free-energy) as a power expansion in $\epsilon$, according to the formula
\begin{multline}\label{eq:Fmseries}
  F_m = - \lim_{\hN \to \infty} \frac{g_m^2}{\hN^2} \log (1 + \epsilon X + \epsilon^2 Y + \dots) 
  =  - \lim_{\hN \to \infty} \frac{g_m^2}{\hN^2} \left[ \epsilon X + \epsilon^2 \left( Y - \frac{X^2}{2} \right) + \dots \right],
\end{multline}
where $X$ and $Y$ come from appropriately collecting terms stemming the various factors in the ``mean value''~\eqref{eq:Z23appendixC}.

Dropping an overall constant, which anyway ``drops out'' when taking the logarithm of $Z$, up to second order in $\epsilon$ we find the following contribution to the ``mean-value''~\eqref{eq:Z23appendixC}
\begin{align}
  & \big\langle  \big\langle \prod_i \cdot \prod_{i\neq j} \big\rangle \big\rangle = \big\langle \big\langle \prod_i \big[1 - 2 \epsilon s^2 \lambda_i^2 + 6 \epsilon^2 s^4 \lambda_i^4 + \mathcal{O}(\epsilon^3) \big] \prod_{i\neq j} \left[ 1 - \frac{\epsilon}{2} A_{ij} +\frac{3}{8} \epsilon^2 A^2_{ij} + \mathcal{O}(\epsilon^3) \right]\big\rangle \big\rangle \nonumber \\
  & {} \hspace{1em} = \big\langle \big\langle \Big[ 1 - 2\epsilon s^2 \sum_i \lambda_i^2 + 6 \epsilon^2 s^4 \sum_i \lambda_i^4 + 2 \epsilon^2 s^4 \sum_i \lambda_i^2 \sum_{k, k \neq i} \lambda_k^2 + \mathcal{O}(\epsilon^3) \Big] \nonumber \\
  & \hspace{1.5em} \times \Big[1 - \frac{\epsilon}{2} \sum_{i \neq j} A_{ij} + \frac{3}{8} \epsilon^2 \sum_{i \neq j} A_{ij}^2 + \frac{\epsilon^2}{8} \sum_{i \neq j} A_{ij}  \!\!\!\!\!\! \sum_{\substack{k \neq l \\ (k,l) \neq (i,j)}}\!\!\!\!\!\!  A_{kl} + \mathcal{O}(\epsilon^3) \Big] \big\rangle \big\rangle \nonumber \\
  & {} \hspace{1em} = \big\langle \big\langle 1 + \epsilon \Big[ - 2 s^2 \sum_i \lambda_i^2 - \um \sum_{i \neq j} A_{ij} \Big]  + \epsilon^2 \Big[ 6 s^4 \sum_i \lambda_i^4 + 2 s^4 \sum_i \lambda_i^2 \! \! \sum_{k, k \neq i} \! \! \lambda_k^2 \nonumber \\
  & \hspace{1.5em} + \frac{3}{8} \sum_{i \neq j} A_{ij}^2 + \frac{1}{8} \sum_{i \neq j} A_{ij}  \!\!\!\!\!\! \sum_{\substack{k \neq l \\ (k,l) \neq (i,j)}}\!\!\!\!\!\!  A_{kl} + s^2 \sum_i \lambda_i^2 \sum_{i \neq j} A_{ij} \Big] + \mathcal{O}(\epsilon^3) \big\rangle \big\rangle,
\end{align}
where we have set $A_{ij}=(\lambda_i^2 + \lambda_j^2 - 2 \lambda_i \lambda_j \cos \beta)$ and introduced the symbol $s$ in place of $\sin \beta /2$. In the rest of the appendix we will also sometime use $C$ in place of $\cos \beta$, so that $C = 1 - 2 s^2$.

To proceed we need to make use of eigenvalues relabeling invariance in order to make use of eigenvalues relabeling invariance to get rid of the various sums in the previous equation~\cite{Siccardi:2009PhD}. Defining
\begin{equation}
  \llangle n_1, \dots, n_k \rrangle \equiv \llangle \lambda_1^{n_1} \dots \lambda_k^{n_k} \rrangle,
\end{equation}
we can write
\begin{align}
  \llangle \sum_i \lambda_i^2 \rrangle = {} & N \llangle \lambda_1^2 \rrangle \equiv N \llangle 2 \rrangle,\\
  \llangle \sum_{i \neq j} A_{ij} \rrangle = {} & N(N-1) \llangle \lambda_1^2 + \lambda_2^2 - 2 \lambda_1 \lambda_2 C \rrangle \nonumber\\
  = {} & 2 N (N - 1) \big( \llangle 2 \rrangle -  \llangle 1,1 \rrangle C),
\end{align}
so that the coefficient of the first order in $\epsilon$ (\textit{i.e.} $X$ in eq.~\eqref{eq:Fmseries}) is given by the expression
\begin{equation}
  -2 s^2 N \llangle 2 \rrangle - N (N - 1) \big( \llangle 2 \rrangle - \llangle 1,1 \rrangle C\big).
\end{equation}
At the next order we have
\begin{align}
  \llangle \sum_i \lambda_i^4 \rrangle = {} & N \llangle 4 \rrangle, \\
  \llangle \sum_i \lambda_i^2 {\sum_j}' \lambda_j^2 \rrangle = {}& N \llangle \lambda_1^2 {\sum_j}' \lambda_j^2 \rrangle = N (N - 1) \llangle \lambda_1^2 \lambda_2^2 \rrangle \nonumber \\
  ={}& N (N - 1) \llangle 2,2 \rrangle,
\end{align}and\begin{align}
  \sum_{i \neq j} A_{ij} {\sum_{k\neq l}}' A_{kl} = {} & N(N - 1) A_{12} {\sum}' A_{ij} = N(N - 1) A_{12} \big[A_{21} + (N - 2) A_{13} \nonumber\\
  & \hspace{.5em} + (N - 2) A_{31} + (N - 2) A_{23} + (N - 2) A_{32} + (N - 2)(N - 3) A_{34} \big] \nonumber\\
  = {} & N(N - 1) \big[ \lambda_1^4 + \lambda_2^4 + 2 \lambda_1^2 \lambda_2^2 + 4 \lambda_1^2 \lambda_2^2 C^2 - 4 \lambda_1^3 \lambda_2 C - 4 \lambda_1 \lambda_2^3 C \nonumber\\
  & \hspace{.5em} + 4 (N - 2)(\lambda_1^4 + \lambda_1^2 \lambda_3^2 - 2 \lambda_1^3 \lambda_3 C + \lambda_1^2 \lambda_2^2 + \lambda_2^2 \lambda_3^2 - 2 \lambda_1 \lambda_2^2 \lambda_3 C - 2 \lambda_1^3 \lambda_3 C \nonumber\\
  & \hspace{.5em} - 2 \lambda_1 \lambda_2 \lambda_3^2 C + 4 \lambda_1^2 \lambda_2 \lambda_3 C^2) + (N-2)(N-3)(\lambda_1^2 \lambda_3^2 +\lambda_2^2 \lambda_3^2 - 2 \lambda_1 \lambda_2 \lambda_3^2 C \nonumber\\
  & \hspace{.5em} + \lambda_1^2 \lambda_4^2 +\lambda_2^2 \lambda_4^2 - 2 \lambda_1 \lambda_2 \lambda_4^2 C - 2 \lambda_1^2 \lambda_3 \lambda_4 C - 2 \lambda_2^2 \lambda_3 \lambda_4 C  + 4 \lambda_1 \lambda_2 \lambda_3 \lambda_4 C^2)\big] \nonumber\\
  = {}& 2 N(N-1)\Big\{ \llangle 4 \rrangle + \llangle 2,2 \rrangle (1+2 C^2) - 4 \llangle 3,1 \rrangle C + 2 (N-2) \big[ \llangle 4 \rrangle \nonumber\\
  & \hspace{.5em} + 3 \llangle 2,2 \rrangle - 4 \llangle 3,1 \rrangle C + 4 \llangle 2,1,1 \rrangle C (C-1)\big] + (N-2)(N-3)\big[2 \llangle 2,2\rrangle \nonumber\\
  & \hspace{.5em} - 4 \llangle 2,1,1\rrangle C + 2\llangle 1,1,1,1 \rrangle C^2 \big]\Big\}.
\end{align}
In the last expressions, we have introduced the shorthand notation
\begin{equation}
  \sum_i \lambda_i^2 {\sum_j}' \lambda_j^2 \equiv \sum_{i=1}^{\hN} \lambda_i^2 {\sum_{\substack{j=1 \\ j\neq i}}^{\hN}} \lambda_j^2 \quad \text{ and } \quad \sum_{i \neq j} A_{ij} {\sum_{k\neq l}}' A_{kl} \equiv  \sum_{i\neq j}^{\hN} A_{ij} \!\!\!\!\!\! \sum_{\substack{k \neq l \\ (k,l) \neq (i,j)}}^{\hN} \!\!\!\!\!\! A_{kl}.
\end{equation}
Lastly we need the formula
\begin{small}\begin{align}
  & \sum \lambda_i^2 \sum A_{ij} = N(N-1) A_{12} \sum \lambda_i^2 = N(N-1) A_{12} [\lambda_1^2 + \lambda_2^2 + (N-2) \lambda_3^2]\nonumber \\
  & {} \hspace{.5em} =  N(N-1) \big[ \lambda_1^4 + \lambda_1^2 \lambda_2^2 + (N-2) \lambda_1^2 \lambda_3^2 + \lambda_1^2 \lambda_2^2 + \lambda_2^4 + (N-2)\lambda_2^2 \lambda_3^2 \nonumber \\
  & \hspace{1.5em} - 2 \lambda_1^3 \lambda_2 C - 2\lambda_1 \lambda_2^3 C - 2(N-2) \lambda_1 \lambda_2 \lambda_3^2 C \big] \nonumber \\
  & {} \hspace{.5em}= 2 N(N-1) \big[ \llangle 4 \rrangle + (N-1) \llangle 2,2 \rrangle - 2 \llangle 3,1 \rrangle C - (N-2) \llangle 2,1,1 \rrangle C\big].
\end{align}\end{small}
The ``mean values'' above are, among many others, available in ref.~\cite{Siccardi:2009PhD} where also details about their derivation are given. Below we list the ones we need here
\begin{align*}
  \llangle 4 \rrangle = {} & 1 + 2 N^2  & \llangle 3 ,1 \rrangle = {} &  1 - 2 N \\
  \llangle 2,2 \rrangle = {} & 1 - N + N^2  & \llangle 2,1 ,1 \rrangle = {} &  2 - N \\
  \llangle 1,1,1,1 \rrangle = {} & 3  &  & \\
  \llangle 2 \rrangle = {} & N  & \llangle 1,1 \rrangle = {} & -1
\end{align*}
Using these results, eq.~\eqref{eq:Fmseries} becomes
\begin{align}\label{eq:FMF}
  F_m^{(2)} =  - \lim_{\hN \to \infty} \frac{g_m^2}{\hN^2} \Big[ &  - 2 \epsilon s^2 \hN^2 - \epsilon N(N-1)(N + C) + \epsilon^2 \frac{\hN^4}{2} (5 + 2C^2) \nonumber \\
  {} & +\epsilon^2 \hN^3 (8 s^4 + 4 s^2 - 2 C^2 + 6 C - 4) \nonumber \\
  {} & + \epsilon^2 \frac{\hN^2}{2} [8 s^4 + 8 s^2 (C - 1) + 8 C^2 - 20 C + 5] \nonumber \\
  {} & + \epsilon^2 \hN (4 s^4 - 4 C s^2 - 3 C^2 + 4 C - 1) \Big],
\end{align}
where, in order to make clear where each term comes from, use was not made of any trigonometric relation. In particular, terms coming from the ``diagonal'' $\prod_i$-product are all proportional to powers of $s^2$. Looking at eq.~\eqref{eq:FMF}, and recalling that $\epsilon$ is inversely proportional to $\hN$, it is possible to appreciate that among the leading terms in $\hN$ there are no contributions coming from the expansion of the $\prod_i$ term. Moreover, if we now enforce the obvious trigonometrical relations, we see that all odd powers of $\hN$ drop out in the $\epsilon^2$ term. The same is true for the even powers in the terms linear in $\epsilon$.

The contribution to the free-energy can thus be written in the form (see eq.~\eqref{eq:Fbetaunostar})
\begin{equation}
  F_m = \mathcal{S}^2 \big[ \frac{h^2 \mathcal{S}}{M_0^3} - \frac{h^4 \mathcal{S}^2}{M_0^6} \frac{5 + 2 \cos^2 \beta}{2} + \mathcal{O}(\mathcal{S}^3)\big].
\end{equation}
Consideration of higher order contributions does not change the conclusion that, in the large-$\hN$ limit, we can neglect the contribution coming from expanding the $\prod_i$-term in the expression for the MM partition function.

\section[$(+,-)$- vs. $(2,3)$-formulation]{$\boldsymbol{(\pm)}$- vs. $\boldsymbol{(2,3)}$-formulation}\label{sec:appA}

In this appendix we want to show that (the MM formulations of) the massive Leigh--Strassler deformations of the $\N=4$ theory, in terms of $\hat{\Phi}_{2,3}$ or $\hat{\Phi}_\pm$ are equivalent in  both the $M_0 \to \infty$ and $M_0 \to 0$ limit.

With respect to the main body of the paper, we will delve in greater detail into computations. Our starting points are the following two formulations of the massive $\beta$d SYM theories in the MM setup
\begin{align}
  Z^{(2,3)} (\beta,M_0) = C_{\hat{N}} \int \du \hat{\Phi}_1 \du \hat{\Phi}_2 \du \hat{\Phi}_3 \exp - \frac{\hat{N}}{g_m} \tr \Big\{ {}& h \hat{\Phi}_1 [\hat{\Phi}_2, \hat{\Phi}_3]_\beta \label{eq:Z23} \\
  {}&+ \sum_{I=2}^3 \frac{M_0}{2} \hat{\Phi}_I^2 +  W\ped{aux}(\hat{\Phi}_1) \Big\}, \nonumber \\
  Z^{(\pm)} (\beta,M_0)= C_{\hat{N}} \int  \du \hat{\Phi}_1 \du \hat{\Phi}_+ \du \hat{\Phi}_-  \exp - \frac{\hat{N}}{g_m} \tr \Big\{ {}& \imath h \hat{\Phi}_1 [\hat{\Phi}_+, \hat{\Phi}_-]_\beta \label{eq:Zpm} \\
  {} &+ M_0 \hat{\Phi}_+ \hat{\Phi}_- +  W\ped{aux}(\hat{\Phi}_1) \Big\}.  \nonumber
\end{align}
We can explicitly verify that these two partition functions indeed correspond to different models for generic values of $\beta$ and $M_0$, by looking at what happens if we perform the substitution
\begin{equation}
  \hat{\Phi}_{(\pm)} = \frac{\hat{\Phi}_2 \pm \imath \hat{\Phi}_3}{\sqrt{2}}\hspace{1em}  \Rightarrow \hspace{1em} \hat{\Phi}_2 = \frac{\hat{\Phi}_+ + \hat{\Phi}_-}{\sqrt{2}}, \hspace{1em} \hat{\Phi}_3 = \frac{\hat{\Phi}_+ - \hat{\Phi}_-}{\sqrt{2}\imath}
\end{equation}
in eq.~\eqref{eq:Z23}. The action, $S^{(2,3)}$ in~\eqref{eq:Z23} becomes
\begin{multline}
  S^{(2,3)} (\hat{\Phi}_1, \hat{\Phi}_{\pm}) = \frac{\hat{N}}{g_m} \tr \{ M_0 \hat{\Phi}_+ \hat{\Phi}_- + \imath h \hat{\Phi}_1 [\hat{\Phi}_+,\hat{\Phi}_-] \cos \beta/2 \\
  + h \hat{\Phi}_1 (\hat{\Phi}_+^2 - \hat{\Phi}_-^2) \sin \beta /2 +  W\ped{aux}(\hat{\Phi_1})\},
\end{multline}
showing that generically the two models are different because
\begin{equation}
  S^{(2,3)} (\hat{\Phi}_1, \hat{\Phi}_{\pm}) \neq S^{(\pm)} (\hat{\Phi}_1, \hat{\Phi}_{\pm}).
\end{equation}
However, in the limits $M_0 \to 0$ and $M_0 \to \infty$, they coincide. To prove this we start by computing the partition function of the $(2,3)$-formulation. Since the action in~\eqref{eq:Z23} is quadratic in any of the three matrices, one can immediately integrate out one of them, say $\hat{\Phi}_2$. Writing the action in the form~\cite{Kawai:2004bz}
\begin{multline}
  S^{(2,3)}_{\beta} (\hat{\Phi}_1, \hat{\Phi}_{2},\hat{\Phi}_{3}) = \frac{\hat{N}}{g_m} \tr \Big\{ \frac{M_0}{2}  \Big( \hat{\Phi}_2 + \frac{h}{M_0} [\hat{\Phi}_3, \hat{\Phi}_1]_\beta \Big)^2\\
  - \frac{h^2}{2 M_0} [\hat{\Phi}_1, \hat{\Phi}_3]_\beta^2  + \frac{M_0}{2} \hat{\Phi}_3^2 + W\ped{aux}(\hat{\Phi}_1) \Big\},
\end{multline}
we get
\begin{align}
  Z^{(2,3)}(\beta,M_0) = C_{\hN} \Big[ \frac{2 \pi g_m}{ \hN M_0} \Big]^{\frac{\hN^2}{2}} \int \du \hat{\Phi}_1 \du \hat{\Phi}_3  {} &  \exp - \frac{\hN}{g_m} \tr \Big\{ \frac{M_0}{2} \hat{\Phi}_3^2 \label{eq:Z23_1} \\
  {} &- \frac{h^2}{2 M_0}  [\hat{\Phi}_1, \hat{\Phi}_3]_\beta^2 + W\ped{aux}(\hat{\Phi}_1)\Big\}. \nonumber
\end{align}
At this point, as usual, we diagonalize $\hat{\Phi}_1$, putting (owing to the ``gauge'' freedom of the MM)
\begin{equation}
  \hat{\Phi}_1 = U
  \left( \! \!
    \begin{array} { ccc }
      \lambda_1 &        & 0  \\
      & \ddots &    \\
      0  &     & \lambda_{\hat{N}} \\
    \end{array}
  \! \! \right) U^{-1} \qquad \qquad \lambda_i \in \mathbb{R}, \: U \,\text{unitary}. \label{eq:CHV}
\end{equation}
With the definition $(U^{-1} \hat{\Phi}_3 U)_{ij} = \mu_{ij} = \mu_{ji}^* =(U^{-1} \hat{\Phi}_3^\dagger U)_{ij}$ we have
\begin{align}
  \tr [\hat{\Phi}_1, \hat{\Phi}_3]_\beta^2 = {} & (\lambda_i \delta_{ij} \mu_{jk} \e^{\imath \beta/2} - \mu_{ij} \lambda_j \delta_{jk} \e^{-\imath \beta/2}) (\lambda_k \delta_{kl} \mu_{li} \e^{\imath \beta/2} - \mu_{kl} \lambda_l \delta_{li} \e^{-\imath \beta/2}) \nonumber \\= {} & -  \sum_{i,k}  \lvert \mu_{ik} \rvert^2 \big(\lambda_i^2 + \lambda_k^2 - 2 \lambda_i \lambda_k \cos \beta \big).
\end{align}
Integration over $\mu_{ik}$ gives
\begin{align}
  & Z^{(2,3)} (\beta, M_0) = C_{\hN} J_{\hN} \left[\frac{2 \pi g_m}{ \hN M_0} \right]^{\hN^2} \int \prod_i \du \lambda_i \prod_i \left[ 1 + \frac{4 h^2}{M_0^2} \lambda_i^2 \sin^2 \beta/2 \right]^{-\frac{1}{2}} \label{eq:Z23appendixA}\\
  &\hspace{.4em} \prod_{i<j} (\lambda_i - \lambda_j)^2 \prod_{i \neq j}  \left\{ 1 + \frac{h^2}{M_0^2} \big[ \lambda_i^2 + \lambda_j^2 - 2 \lambda_i \lambda_j \cos \beta \big] \right\}^{-\frac{1}{2}} \e^{- \frac{\hN}{g_m} \sum_i W\ped{aux}(\lambda_i)} ,\nonumber
\end{align}
where $ J_{\hN}$ has already been defined,  while the factor $\prod_{i < j} (\lambda_i - \lambda_j)^2$ is the usual Jacobian associated with the change of variables~(\ref{eq:CHV}).

Despite the presence of the $M_0^{-\hN^2}$ factor in front of the integral, $Z^{(2,3)}$ has a finite $M_0 \to 0$ limit, as can be seen by suitably distributing the powers of $M_0$ among the various terms. Actually, keeping only the leading contribution (as $\hN\to\infty$, see appendix~\ref{app:subleading}), in the  $M_0\to 0$ limit $Z^{(2,3)}$ becomes
\begin{multline}\label{eq:Z23_f}
  Z^{(2,3)}(\beta,0) = C_{\hN} J \left[\frac{2 \pi g_m}{\hN h}\right]^{\hN^2} \int \prod_i \du \lambda_i \prod_{i < j} \frac{(\lambda_i - \lambda_j)^2}{ \lambda_i^2 + \lambda_j^2 - 2 \lambda_i \lambda_j \cos \beta} \\
  \exp\{- \frac{\hN}{g_m} \sum_i W\ped{aux}(\lambda_i)\}.
\end{multline}
If we instead start from eq.~\eqref{eq:Zpm}, we can rewrite $S^{(\pm)}_{\beta}(\hat{\Phi}_1, \hat{\Phi}_{\pm})$ in the form
\begin{align}
  S^{(\pm)}_{\beta}(\hat{\Phi}_1, \hat{\Phi}_{\pm}) = \frac{\hat{N}}{g_m} \tr \left\{ W\ped{aux}(\hat{\Phi}_1) + M_0 \hat{\Phi}_+ \left[ \one \otimes \one + \frac{\imath h}{M_0} \big(\e^{\imath \beta /2 } \hat{\Phi}_1 \otimes \one - \one \otimes \hat{\Phi}_1 \e^{-\imath \beta/2} \big) \right] \hat{\Phi}_- \right\} .\nonumber
\end{align}
After integrating out simultaneously the two conjugate matrices $\Phi_{\pm}$, the resulting partition function is
\begin{align}
  Z^{(\pm)} (\beta, M_0) = C_{\hat{N}}\left[\frac{ 2 \pi g_m}{\hat{N} M_0} \right]^{\hat{N}^2} \int \du \hat{\Phi}_1 \frac{\e^{-\frac{\hN}{g_m} \tr W\ped{aux}(\Phi)}}{\text{det} \Big[ \one \otimes \one + \frac{ \imath h}{M_0} \big( \e^{\imath \beta /2 } \hat{\Phi}_1 \otimes \one - \one \otimes \hat{\Phi}_1 \e^{-\imath \beta/2} \big) \Big]},\nonumber
\end{align}
which, diagonalising $\hat{\Phi}_1$, can be rewritten as
\begin{multline}
  Z^{(\pm)} (\beta,M_0) = C_{\hat{N}} J_{\hN} \left[\frac{ 2 \pi g_m}{\hat{N} M_0} \right]^{\hat{N}^2} \int \prod_i \du \lambda_i \prod_{i<j}(\lambda_i - \lambda_j)^2 \e^{-\frac{\hN}{g_m} \sum W\ped{aux}(\lambda_i)} \nonumber \\
  \prod_{i,j} \frac{1}{1 + \frac{\imath h}{M_0} (\lambda_i \e^{\imath \beta/2} - \lambda_j \e^{-\imath \beta/2})} \nonumber \\
  = C_{\hat{N}} J_{\hN} \left[ \frac{ 2 \pi g_m}{\hat{N} M_0} \right]^{\hat{N}^2} \int \prod_i \du \lambda_i  \prod_{i<j} (\lambda_i - \lambda_j)^2 \prod_i \frac{1}{1 + \frac{\imath h}{M_0} \lambda_i (\e^{\imath \beta/2} - \e^{- \imath \beta /2})} \nonumber \\
  \prod_{i \neq j} \frac{1}{\frac{M_0}{h} + \imath (\lambda_i \e^{\imath \beta/2} - \lambda_j \e^{-\imath \beta/2})} \e^{-\frac{\hN}{g_m} \sum W\ped{aux}(\lambda_i)}.
\end{multline}
Again, redistributing the $M_0$ dependence, we may arrange the last expression in the form
\begin{align}\label{ZPM}
  & Z^{(\pm)}(\beta,M_0) = C_{\hat{N}} J_{\hN} \left[\frac{ 2 \pi g_m}{\hat{N} h} \right]^{\hat{N}^2} \int \prod_i \du \lambda_i \prod_{i<j} (\lambda_i -\lambda_j)^2 \\
  & \hspace{.5em} \prod_i \frac{ 1}{\frac{M_0}{h} + \imath \lambda_i (\e^{\imath \beta/2} - \e^{-\imath \beta/2})} \prod_{i \neq j} \frac{1}{\frac{M_0}{h} + \imath (\lambda_i \e^{\imath \beta/2} - \lambda_j \e^{-\imath \beta/2})}\e^{-\frac{\hN}{g_m} \sum W\ped{aux}(\lambda_i)}. \nonumber
\end{align}
Repeating the arguments given in appendix~\ref{app:subleading}, and already exploited to deal with the (2,3)-formulation, one can show that the diagonal product in the second line does not contribute to the large-$\hat{N}$ limit. Then, in the $M_0 \to 0$ limit one gets
\begin{multline}\label{eq:ZPM_f}
  Z^{(\pm)}(\beta,M_0) = C_{\hat{N}} J_{\hN} \left[\frac{ 2 \pi g_m}{\hat{N} h} \right]^{\hat{N}^2} \int \prod_i \du \lambda_i \prod_{i < j} \frac{(\lambda_i - \lambda_j)^2}{\lambda_i^2 + \lambda_j^2 - 2 \lambda_i \lambda_j \cos \beta} \\
  \exp\left\{-\frac{\hN}{g_m} \sum W\ped{aux}(\lambda_i)\right\},
\end{multline}
which coincides with eq.~\eqref{eq:Z23_f}, as announced.

We now move to the $M_0\to \infty$ case. In this limit obviously only the mass terms need to be kept in the potential. It is then clear that the two zero-dimensional functional integrals are just gaussian and identical.

Actually, the latter is quite an expected result, since, in the dual gauge theory description, it is known that in the $M_0 \to \infty$ limit we decouple completely the $\Phi_{2,3}$ degrees of freedom in one case and the $\Phi_{\pm}$ degrees of freedom in the other. What is left is simply an integral over the last chiral field that is not affected by the rotation. In other words, there is no difference between the two possible forms of $Z(\beta,M_0)$ in the $M_0 \to \infty$ limit.

\providecommand{\href}[2]{#2}\begingroup\raggedright

\end{document}